\documentclass[a4paper]{article}
\usepackage{INTERSPEECH2020}
\usepackage{multirow, multicol}
\usepackage{booktabs}
\usepackage{graphicx}
\usepackage{amssymb}
\usepackage[table]{xcolor}
\usepackage{endnotes}
\usepackage{hyperref}
\hypersetup{
}
\usepackage{dblfloatfix}
\usepackage{adjustbox}
\usepackage{enumitem}
\usepackage{verbatim}
\usepackage{caption}

\usepackage{threeparttable, tablefootnote}

\title{Robust Text-Dependent Speaker Verification via Character-Level Information Preservation for the SdSV Challenge 2020}

\name{Sung Hwan Mun, Woo Hyun Kang, Min Hyun Han and Nam Soo Kim}
\address{
 Department of Electrical and Computer Engineering and the Institute of New Media and Communications, Seoul National University, Seoul, South Korea}
\email{\{shmun, whkang, mhhan\}@hi.snu.ac.kr, nkim@snu.ac.kr}

\begin{document}

\maketitle
\begin{abstract}
This paper describes our submission to Task 1 of the Short-duration Speaker Verification (SdSV) challenge 2020. Task 1 is a text-dependent speaker verification task, where both the speaker and phrase are required to be verified. The submitted systems were composed of TDNN-based and ResNet-based front-end architectures, in which the frame-level features were aggregated with various pooling methods (e.g., statistical, self-attentive, ghostVLAD pooling). Although the conventional pooling methods provide embeddings with a sufficient amount of speaker-dependent information, our experiments show that these embeddings often lack phrase-dependent information. To mitigate this problem, we propose a new pooling and score compensation methods that leverage a CTC-based automatic speech recognition (ASR) model for taking the lexical content into account. Both methods showed improvement over the conventional techniques, and the best performance was achieved by fusing all the experimented systems, which showed 0.0785\% MinDCF and 2.23\% EER on the challenge's evaluation subset.
\end{abstract}
\section{Introduction}
This paper presents our submission to Task 1 of the Short-duration Speaker Verification (SdSV) Challenge 2020. The main purpose of this challenge is to evaluate new techniques for speaker verification in a short duration scenario \cite{sdsvc2020plan}. The evaluation dataset used for the SdSV Challenge 2020 was derived from the multi-purpose DeepMine dataset \cite{zeinali2018deepmine, zeinali2019multi}. We submitted the systems to Task 1 of the SdSV Challenge 2020, which was focused on text-dependent speaker verification (TD-SV).

During the past decade, there has been significant improvement in the field of text-independent speaker verification (TI-SV), which is mainly attributed to the development of deep neural network (DNN) based speaker embedding. These mechanisms have been developed through a variety of methods such as deeper architectures \cite{he2016deep, chung2019delving, snyder2018xvector, snyder2019etdnn}, pooling strategies \cite{snyder2018xvector, zhu2018asp, xie2019gvlad, han2020minep}, and different objective functions \cite{liu2017asoft, wang2018amsoft, wang2018amsoft2, deng2019aamsoft, chung2020in, wan2018ge2e}. However, these techniques are optimized to discriminate only the speaker and may be ineffective in the TD-SV task in which the lexical context, as well as the speaker, is considered \cite{zeinali2017hmm}.

On the other hand, there have also been various efforts to boost performance in the TD-SV. In \cite{larcher2014rsr}, Larcher et al. used an HMM-based system named HiLAM to model each speaker and each senone state. H. Zeinali et al. \cite{zeinali2017hmm} proposed a straightforward HMM-based extension of the i-vector approach \cite{dehak2010ivector}, which allows i-vectors to contain sufficient text-dependent information. In \cite{lei2014novel}, Y. Lei et al. used DNN to estimate the posteriors of the frames for calculation of sufficient statistics and E. Variani et al. \cite{variani2014dvector} extracted frame-level representations termed d-vector through a hidden layer of DNN for the TD-SV task. In \cite{matvejka2016analysis}, Matejka et al. employed bottle-neck DNN features concatenated to other acoustic features to improve the performance, and Zhang et al. \cite{zhang2017end} proposed an attention aggregation-based end-to-end TD-SV system which takes the speaker and phonetic information into account.

In this paper, we focus on preserving the character-level information. Overall, our contributions are as follows:
\begin{itemize}[topsep=0.3pt, itemsep=0.3pt, partopsep=0.3pt, parsep=0.3pt]
  \item \textbf{Character-level pooling:} We introduce the aggregation method using the estimated frame-level posterior obtained from an automatic speech recognition (ASR) model. Experiments show that this method is valid and effective in the TD-SV through the results on the challenge's progress and evaluation subsets.
  \item \textbf{Score compensation:} We propose the score compensation method where the probability of pass-phrase is estimated. Our experiments show that the usage of the proposed score compensation significantly enhances performance in the TD-SV even if the embedding network is trained only to classify the speaker.
\end{itemize}
Furthermore, the fusion of different systems we employed produces the best performance on the challenge’s trial subsets.

The rest of this paper is organized as follows: Section 2 describes all components of our systems and Section 3 presents the experimental conditions and results on the challenge's trial subsets we submitted. Finally, we conclude in section 4.


\section{System components description}
\subsection{Front-end}
In our systems, we used two types of front-end networks: TDNN-based \cite{snyder2018xvector} and ResNet-based \cite{chung2019delving} architectures.

\vspace{2.0mm}\noindent\textbf{TDNN-based architecture}.
The configuration of TDNN-based systems is shown in Table 1. The usage for each system is described in Section 2.2 and 2.6. The input acoustic feature used in this architectures was log Mel-filterbank energies calculated from 20ms windows with a 10ms hop size and extracted via the Librosa toolkit \cite{mcfee2015librosa}. We selected 512 and 580 speaker embedding dimensions for statistics pooling and character-level pooling respectively. Our implementation and speaker embedding network training was done using Tensorflow toolkit \cite{abadi2016tensorflow}.


\begin{table*}[ht]
  \caption{TDNN-based front-end configuration for character-level pooling and score compensation. \((d\times n)\) indicates concatenation of n vectors, where the dimensionality of each vector is d. T: The number of segment frames, N: The number of speakers, M: The number of phrase types, CLP: Character-Level Pooling, LC: Locally-Connected, FC: Fully-Connected, BN: Batch Normalization.}
  \label{tdnn}
  \centering
  \renewcommand{\arraystretch}{1.05}
  \renewcommand{\tabcolsep}{2.28mm}
  {\footnotesize
  \begin{tabular}{lllllll}
    \toprule
    \multirow{2}{*}{\textbf{Layer}} &
    \multicolumn{3}{l}{\textbf{Configuration for character-level pooling method}} &
    \multicolumn{3}{l}{\textbf{Configuration for score compensation method}} \\
    \cmidrule(lr){2-4}  \cmidrule(lr){5-7} 
            &
    \textbf{TDNN} &
    \textbf{Context} &
    \textbf{Output Size} &
    \textbf{TDNN} &
    \textbf{Context} &
    \textbf{Output Size} \\

    \midrule
    \multicolumn{1}{l|}{Input} &
    Log Mel-FBANK &
    -- &
    \multicolumn{1}{l||}{$\text{64} \times \textit{T}$} &
    Log Mel-FBANK &
    -- &
    $\text{64} \times \textit{T}$ \\
    
    \arrayrulecolor{black!30}\specialrule{.01em}{.0em}{.0em}\arrayrulecolor{black}  
    \multicolumn{1}{l|}{Frame1} &
    \text{512}, \text{stride 2, ReLU, BN} &
    \text{5}, \text{[\textit{t}-2 : \textit{t}+2]} &
    \multicolumn{1}{l||}{$\text{512} \times \textit{T}$} &
    \text{1536}, \text{stride 2, ReLU, BN} &
    \text{5}, \text{[\textit{t}-2 : \textit{t}+2]} &
    $\text{1536} \times \textit{T}$ \\
    
    \multicolumn{1}{l|}{Frame2} &
    \text{512}, \text{stride 1, ReLU, BN} &
    \text{3}, \text{[\textit{t}-2, \textit{t}, \textit{t}+2]} &
    \multicolumn{1}{l||}{$\text{512} \times \textit{T}$} &
    \text{512}, \text{stride 1, ReLU, BN} &
    \text{3}, \text{[\textit{t}-2, \textit{t}, \textit{t}+2]} &
    $\text{512} \times \textit{T}$ \\
    
    \multicolumn{1}{l|}{Frame3} &
    \text{512}, \text{stride 1, ReLU, BN} &
    \text{3}, \text{[\textit{t}-3, \textit{t}, \textit{t}+3]} &
    \multicolumn{1}{l||}{$\text{512} \times \textit{T}$} &
    \text{512}, \text{stride 1, ReLU, BN} &
    \text{3}, \text{[\textit{t}-3, \textit{t}, \textit{t}+3]} &
    $\text{512} \times \textit{T}$ \\
    
    \multicolumn{1}{l|}{Frame4} &
    \text{512}, \text{stride 1, ReLU, BN} &
    \text{1}, \text{[\textit{t}]} &
    \multicolumn{1}{l||}{$\text{512} \times \textit{T}$} &
    \text{256}, \text{stride 1, ReLU, BN} &
    \text{1}, \text{[\textit{t}]} &
    $\text{256} \times \textit{T}$ \\
    
    \multicolumn{1}{l|}{Frame5} &
    \text{1536}, \text{stride 1, ReLU, BN} &
    \text{1}, \text{[\textit{t}]} &
    \multicolumn{1}{l||}{$\text{1536} \times \textit{T}$} &
    \text{256}, \text{stride 1, ReLU, BN} &
    \text{1}, \text{[\textit{t}]} &
    $\text{256} \times \textit{T}$ \\
    
    \arrayrulecolor{black!30}\specialrule{.01em}{.0em}{.0em}\arrayrulecolor{black} 
    \multicolumn{1}{l|}{Pooling} &
    \text{CLP} &
    \textit{T}, \text{[1 : \textit{T}]} &
    \multicolumn{1}{l||}{($\text{1536} \times \text{29}$) $\times$ $\text{1}$} &
    \text{CLP} &
    \textit{T}, \text{[1 : \textit{T}]} &
    ($\text{256} \times \text{29}$) $\times$ $\text{1}$ \\
   
    \arrayrulecolor{black!30}\specialrule{.01em}{.0em}{.0em}\arrayrulecolor{black} 
    \multicolumn{1}{l|}{Segment1} &
    \text{LC \, \textit{(speaker embedding)}} &
    \textit{T}, \text{[1 : \textit{T}]} &
    \multicolumn{1}{l||}{($\text{20} \times \text{29}$) $\times$ $\text{1}$} &
    \text{LC} &
    \textit{T}, \text{[1 : \textit{T}]} &
    ($\text{20}$ $\times$ $\text{29}$) $\times$ $\text{1}$ \\
    
    \multicolumn{1}{l|}{Segment2} &
    \text{FC} &
    \textit{T}, \text{[1 : \textit{T}]} &
    \multicolumn{1}{l||}{$\text{512}$ $\times$ $\text{1}$} &
    \text{FC} &
    \textit{T}, \text{[1 : \textit{T}]} &
    $\text{512}$ $\times$ $\text{1}$ \\
    
    \arrayrulecolor{black!30}\specialrule{.01em}{.0em}{.0em}\arrayrulecolor{black}     
    \multicolumn{1}{l|}{Softmax} &
    \text{FC} &
    \textit{T}, \text{[1 : \textit{T}]} &
    \multicolumn{1}{l||}{$\emph{N}$ $\times$ $\text{1}$} &
    \text{FC \, \textit{(posterior of phrase)}} &
    \textit{T}, \text{[1 : \textit{T}]} &
    $\emph{M}$ $\times$ $\text{1}$ \\
    
    \bottomrule
  \end{tabular}
  }
\end{table*}


\begin{table}[ht]
  \caption{Thin ResNet34-based front-end configuration. All convolutional layers have 1 zero-padding.}
  \label{resnet}
  \centering
  \renewcommand{\arraystretch}{1.05}
  \renewcommand{\tabcolsep}{3.65mm}
  {\footnotesize
  \begin{tabular}{lll}
    \toprule
    \multicolumn{1}{l}{\textbf{Layer}} &
    \textbf{Thin ResNet34} &
    \textbf{Output Size} \\
    \midrule
    \multicolumn{1}{l|}{Input} &
    \text{STFT} &
    \text{257} $\times$ \textit{T} $\times$ \text{1} \\
    \arrayrulecolor{black!30}\specialrule{.01em}{.0em}{.0em}\arrayrulecolor{black}
    \multicolumn{1}{l|}{\multirow{2}{*}{Conv1}} &
    \text{7}$\times$\text{7}, \text{16}, \text{stride 2} &
    \text{129} $\times$ \textit{T}\text{\,/\,2} $\times$ \text{16} \\
    \multicolumn{1}{l|}{} &
    \text{3}$\times$\text{3}, \text{max pooling}, \text{stride 2} &
    \text{65} $\times$ \textit{T}\text{\,/\,4} $\times$ \text{16} \\
    \multicolumn{1}{l|}{Conv2} &
    $\begin{bmatrix}\text{3}\times\text{3}, \>\text{16} \\ 
    \text{3}\times\text{3}, \>\text{16}\end{bmatrix}$$\times$\text{3}, \text{stride 1} &
    \text{65} $\times$ \textit{T}\text{\,/\,4} $\times$ \text{16} \\
    \multicolumn{1}{l|}{Conv3} &
    $\begin{bmatrix}\text{3}\times\text{3}, \>\text{32} \\ 
    \text{3}\times\text{3}, \>\text{32}\end{bmatrix}$$\times$\text{4}, \text{stride 2} &
    \text{33} $\times$ \textit{T}\text{\,/\,8} $\times$ \text{32} \\
    \multicolumn{1}{l|}{Conv4} &
    $\begin{bmatrix}\text{3}\times\text{3}, \>\text{64} \\ 
    \text{3}\times\text{3}, \>\text{64}\end{bmatrix}$$\times$\text{6}, \text{stride 2}  &
    \text{17} $\times$ \textit{T}\text{\,/\,16} $\times$ \text{64} \\
    \multicolumn{1}{l|}{Conv5} & 
    $\begin{bmatrix}\text{3}\times\text{3}, \>\text{128} \\ 
    \text{3}\times\text{3}, \>\text{128}\end{bmatrix}$$\times$\text{3}, \text{stride 2} &
    \text{9} $\times$ \textit{T}\text{\,/\,32} $\times$ \text{128} \\
    \multicolumn{1}{l|}{FC} &
    \text{9}$\times$ \text{1}, \text{512}, \text{stride 1} &
    \text{1}$\times$ \textit{T}\text{\,/\,32} $\times$ \text{512} \\
    \bottomrule
  \end{tabular}
  }
  \end{table}


\vspace{2.0mm}\noindent\textbf{ResNet-based architecture}.
We used Thin ResNet34 architecture recently proposed in \cite{chung2019delving} (Table 2). Compared to the original ResNet \cite{he2016deep}, it has only a quarter of channels in each residual block. In this architecture, we used 257-dimensional short-time Fourier transform (STFT) with 200-300 frames crop as input acoustic feature and chose 512 dimensions for embedding. For implementation, we used Pytorch toolkit \cite{paszke2019pytorch} and developed the systems based on the architectures in \cite{chung2020in}.

\subsection{Pooling methods}
In the TI-SV task, various pooling mechanisms have been proposed such as statistics pooling \cite{snyder2018xvector}, self-attentive pooling \cite{zhu2018asp}, learnable dictionary encoding (LDE) pooling \cite{xie2019gvlad}, mutual information neural estimate (MINE) based pooling \cite{han2020minep}. In this work, we employed a variety of pooling methods proposed in the TI-SV task. On top of that, we propose a pooling strategy suitable for the TD-SV, i.e., character-level pooling, which leverages a frame-level probability distribution of each character estimated from the end-to-end based ASR model. The pooling methods we used are as follows:
\begin{itemize}
\item Statistics Pooling (SP) \cite{snyder2018xvector}
\item Self-Attentive Pooling (SAP) \cite{zhu2018asp}
\item GhostVLAD Pooling (GVP) \cite{xie2019gvlad}
\item Character-Level Pooling (CLP)
\end{itemize}
\vspace{2.0mm}\noindent\textbf{Character-level pooling.}
To extract the utterance-level representation appropriate for the TD-SV, we exploit the character posterior probabilities of each frame-level feature. The probability of a character given a frame-level feature, i.e., the posterior, is denoted by:
\begin{equation}
  \pi_{k,i} = P(C=c_k \vert \vec{h_i})
  \label{eq1}
\end{equation}
where the set \(C = \{c_k\vert\>c_k \> is \> k^{th} character,\, 1\leq k \leq K\}\),  and \(\vec{h_i}\) is the \(i^{th} \) frame-level feature with \(D_1\) dimensions where \(1 \leq i \leq T\). \(K\) indicates the number of symbols in the character set, and \(T\) is the number of segment frames. To estimate \(\pi_{k,i}\), we leveraged the decoder outputs of the end-to-end based ASR model, termed \(Jasper\), proposed by \cite{li2019jasper}. Since this model was trained by using the Connectionist Temporal Classification (CTC) loss, our character set consisted of a total of 29 symbols including all alphabets (a-z), the space symbol, the apostrophe symbol and the blank symbol used by the CTC loss. Then, the aggregation for character-level representation is as follows:
\begin{equation}
  \vec{v_k} = {{\sum_{i=1}^{T}\pi_{k,i}\vec{h_i} + \tau}  \over {\sum_{i=1}^{T}\pi_{k,i} + \tau}}
  \label{eq2}
\end{equation}
\begin{equation}
  \vec{v} = \begin{pmatrix} \vec{v_1}^T \,\,\vert\,\,\, \,\ldots\, \,\,\,\vert\,\,\vec{v_K}^T \end{pmatrix}^T
  \label{eq3}
\end{equation}
where \(\tau\) is a constant added to avoid divergence. All the character-level representations are concatenated as \(\vec{v}\), and then it's passed through the locally-connected layer, which has \(K\)-part fully-connected layers for reducing dimensions and taking character-level affine transformation.
\begin{equation}
  \vec{e_k} = f(\vec{W_k}\vec{v_k}+\vec{b_k})
  \label{eq4}
\end{equation}
\begin{equation}
  \vec{e} = \begin{pmatrix} \vec{e_1}^T \,\,\vert\,\,\, \,\ldots\, \,\,\,\vert\,\,\vec{e_K}^T \end{pmatrix}^T
  \label{eq5}
\end{equation}
Where \(\vec{W_k}\) and \(\vec{b_k}\) indicate trainable parameters with \(D_2\times D_1\) and \(D_2\) dimensions respectively and \(f(\cdot)\) means a non-linear activation function. Finally, we can obtain an utterance-level embedding \(\vec{e}\) (See Table 1).
%
%
\subsection{Objective functions}
In our work, we made use of various objective functions conventionally used in speaker embedding training. Some variants of softmax-based classification loss were employed in our systems. Also, we used end-to-end based losses which directly optimize distance metrics such as Euclidean or Cosine distance. The objective functions used in the systems are as follows:
\begin{itemize}
\item Standard Softmax
\item Additive Margin Softmax (AM-Softmax) \cite{wang2018amsoft, wang2018amsoft2}
\item Additive Angular Margin Softmax (AAM-Softmax) \cite{deng2019aamsoft}
\item Angular Prototypical Loss (A-Prototypical) \cite{chung2020in}
\item Generalized End-to-End Loss (GE2E) \cite{wan2018ge2e}
\end{itemize}
%
%
\begin{table*}[t]
\centering
\caption{Results on the Trial Subsets for the SdSV Challenge 2020 \textbf{without AS-Norm \& Score Compensation}. TDT: Text-Dependent Training, Deep: DeepMine, Vox1: VoxCeleb1, Vox2: VoxCeleb2, Libri: LibriSpeech.}
\label{tab:my-table}
\resizebox{\textwidth}{!}{
{\scriptsize
\begin{tabular}{lllllllll}
\toprule
\multirow{2}{*}{\#} & \multirow{2}{*}{\textbf{Front-End}} & \multirow{2}{*}{\textbf{Objectives}} & \multirow{2}{*}{\textbf{Pooling}} & \multirow{2}{*}{\textbf{Training Dataset}} & \multicolumn{2}{l}{\textbf{Progress subset}} & \multicolumn{2}{l}{\textbf{Evaluation subset}} \\ \cmidrule(lr){6-7}  \cmidrule(lr){8-9}
 &  &  &  &  & \textbf{MinDCF} & \textbf{EER{[}\%{]}} & \textbf{MinDCF} & \textbf{EER{[}\%{]}} \\ 
\midrule
1 & \multirow{3}{*}{\textbf{TDNN}} & \multirow{3}{*}{\textbf{Softmax}} & \multirow{3}{*}{\textbf{CLP}} & Deep & 0.3755 & 9.19 & 0.3775 & 9.18 \\
\textbf{2} &  &  &  & \textbf{Deep / Vox1} & 0.3571 & \textbf{8.45} & 0.3585 & \textbf{8.48} \\
3 &  &  &  & Deep / Vox1 / Vox2 & 0.4044 & 8.97 & 0.4066 & 9.00 \\
\addlinespace[.3em]
\textbf{4} & \textbf{TDNN} & \textbf{Softmax (TDT)} & \textbf{CLP} & \textbf{Deep} & \textbf{0.3547} & 8.82 & \textbf{0.3554} & 8.88 \\
\addlinespace[.3em]
5 & \multirow{3}{*}{TDNN} & \multirow{3}{*}{Softmax} & \multirow{3}{*}{SP} & Deep & 0.8679 & 17.18 & 0.8688 & 17.25 \\
6 &  &  &  & Deep / Vox1 & 0.7636 & 14.37 & 0.7641 & 14.45 \\
7 &  &  &  & Deep / Vox1 / Vox2 & 0.6511 & 12.71 & 0.6539 & 12.77 \\
\addlinespace[.3em]
8 & ResNet34 & Softmax & GVP & Vox2 & 0.8891 & 14.84 & 0.8897 & 14.87 \\
\addlinespace[.3em]
9 & \multirow{2}{*}{ResNet34} & \multirow{2}{*}{AAM-Softmax} & \multirow{2}{*}{SAP} & Deep / Vox1 / Vox2 & 0.9030 & 16.09 & 0.9021 & 16.12 \\
10 &  &  &  & Deep / Vox1 / Vox2 / Libri & 0.9157 & 16.39 & 0.9159 & 16.45 \\
\addlinespace[.3em]
11 & \multirow{2}{*}{ResNet34} & \multirow{2}{*}{AM-Softmax} & \multirow{2}{*}{SAP} & Deep / Vox1 / Vox2 & 0.8944 & 15.76 & 0.8931 & 15.84 \\
12 &  &  &  & Deep / Vox1 / Vox2 / Libri & 0.9195 & 16.42 & 0.9181 & 16.47 \\
\addlinespace[.3em]
13 & \multirow{2}{*}{ResNet34} & \multirow{2}{*}{A-Prototypical} & \multirow{2}{*}{SAP} & Vox2 & 0.7957 & 13.35 & 0.7973 & 13.33 \\
14 &  &  &  & Deep / Vox1 / Vox2 & 0.8659 & 15.92 & 0.8652 & 15.98 \\
\addlinespace[.3em]
15 & ResNet34 & GE2E & SAP & Deep / Vox1 / Vox2 & 0.9226 & 16.39 & 0.9212 & 16.45 \\

\arrayrulecolor{black!30}\specialrule{.01em}{.15em}{.15em}
16 & \multicolumn{4}{c}{x-vector baseline \textit{(provided by SdSV)}} & 0.5290 & 9.05 & 0.5287 & 9.05 \\
\arrayrulecolor{black}\bottomrule
\end{tabular}%
}}
\end{table*}

%
\subsection{Back-end}
In the back-end module, we only used cosine similarity as a scoring method between the two speaker embeddings. No Linear Discriminant Analysis (LDA), Within-Class Covariance Normalization (WCCN), and Probabilistic Linear Discriminant Analysis (PLDA) was applied in this work. 
%
%
\subsection{Score normalization}
To minimize the domain mismatch (e.g., languages, recording environments, etc.) between the training and the evaluation set and to normalize the distribution of scores during fusion between different models, we used the Adaptive Symmetric Score Normalization (AS-Norm) \cite{matejka2017analysis}. We selected speaker-phrase dependent models in DeepMine Task1 Train Partition (i.e., in-domain training data) as cohort set and used the most similar top 300 scoring files to calculate normalization variables of enrollment and test sets respectively.
%
%
\subsection{Score compensation}
Since the TI-SV approaches focus on minimizing the within-speaker variability, the embedding vectors may lack crucial information on the lexical content. Thus such embedding vectors are improper for the TD-SV experiments. To complement the lost contextual discriminability, we introduce a score compensation method. Firstly, we define the posterior of the phrase as follows:
\begin{equation}
  \vec{u_{X}} = \begin{pmatrix} \, P(U=u_1 \vert \vec{X}) \,,\,\,\, \ldots \,\,\,\,,\, P(U=u_M\vert\vec{X}) \,\, \end{pmatrix}^T
  \label{eq1}
\end{equation}
\begin{equation}
  \sum_{j=1}^{M}p(U=u_j\vert\vec{X}) = 1
  \label{eq1}
\end{equation}
where the set \(U = \{u_j\vert\>u_j \> is \> j^{th} phrase,\, 1\leq j \leq M\}\), \(M\) is the number of phrase types in the TD-SV's dataset, \(X\) is an acoustic feature such as MFCC. We estimate the posterior \( p(U=u_j \vert X) \) using softmax layers of TDNN-based network. The architecture of this network is identical to the configuration of TDNN-based architecture for CLP but composed of smaller size layers to prevent overfitting (See Table 1). For training the network, we used DeepMine Task1 Train Partition which includes 10 types of phrases. Finally, we compute the compensation factor and the total score between \(X\) and \(Y\) as follows:
\begin{equation}
  s_{X,Y}^{phr} = \vec{u_{X}}^T\,\vec{u_{Y}}
  \label{eq8}
\end{equation}
\begin{equation}
  s_{X,Y} = \tilde{s}_{X,Y}^{spk} + \alpha s_{X,Y}^{phr}
  \label{eq9}
\end{equation}
where \(s_{X,Y}^{phr}\) is compensation factor, \(\tilde{s}_{X,Y}^{spk}\) is the normalized (AS-Norm) score between embeddings of X and Y, \(\alpha\) is a scale factor, and \(s_{X,Y}\) is the total score between X and Y.
%
%
\begin{table*}[!ht]
\centering
\caption{Results on the Trial Subsets \textbf{with AS-Norm \& Score Compensation and the Fusions}. {\dag}: Fusion is equal-weighted sum.}
\label{tab:my-table}
\resizebox{\textwidth}{!}{
{\scriptsize
\begin{tabular}{lllllllll}
\toprule
\multirow{2}{*}{\#} & \multirow{2}{*}{\textbf{Front-End}} & \multirow{2}{*}{\textbf{Objectives}} & \multirow{2}{*}{\textbf{Pooling}} & \multirow{2}{*}{\textbf{Training Dataset}} & \multicolumn{2}{l}{\textbf{Progress subset}} & \multicolumn{2}{l}{\textbf{Evaluation subset}} \\
\cmidrule(lr){6-7}  \cmidrule(lr){8-9}
 &  &  &  &  & \textbf{MinDCF} & \textbf{EER{[}\%{]}} & \textbf{MinDCF} & \textbf{EER{[}\%{]}} \\
\midrule
1 & \multirow{3}{*}{TDNN} & \multirow{3}{*}{Softmax} & \multirow{3}{*}{CLP} & Deep & 0.2164 & 5.79 & 0.2185 & 5.82 \\
2 &  &  &  & Deep / Vox1 & 0.1845 & 4.72 & 0.1856 & 4.80 \\
3 &  &  &  & Deep / Vox1 / Vox2 & 0.1892 & 4.91 & 0.1918 & 4.98 \\
\addlinespace[.3em]
4 & TDNN & Softmax (TDT) & CLP & Deep & 0.2327 & 5.88 & 0.2333 & 5.98 \\
\addlinespace[.3em]
5 & \multirow{3}{*}{TDNN} & \multirow{3}{*}{Softmax} & \multirow{3}{*}{SP} & Deep & 0.2540 & 7.35 & 0.2554 & 7.42 \\
6 &  &  &  & Deep / Vox1 & 0.2069 & 5.54 & 0.2085 & 5.63 \\
7 &  &  &  & Deep / Vox1 / Vox2 & 0.1730 & 4.49 & 0.1753 & 4.55 \\
\addlinespace[.3em]
8 & ResNet34 & Softmax & GVP & Vox2 & 0.1993 & 4.59 & 0.2017 & 4.65 \\
\addlinespace[.3em]
9 & \multirow{2}{*}{ResNet34} & \multirow{2}{*}{AAM-Softmax} & \multirow{2}{*}{SAP} & Deep / Vox1 / Vox2 & 0.1327 & 3.15 & 0.1332 & 3.21 \\
10 &  &  &  & Deep / Vox1 / Vox2 / Libri & 0.1321 & 3.30 & 0.1325 & 3.33 \\
\addlinespace[.3em]
\textbf{11} & \multirow{2}{*}{\textbf{ResNet34}} & \multirow{2}{*}{\textbf{AM-Softmax}} & \multirow{2}{*}{\textbf{SAP}} & \textbf{Deep / Vox1 / Vox2} & \textbf{0.1299} & \textbf{3.13} & \textbf{0.1307} & \textbf{3.18} \\
12 &  &  &  & Deep / Vox1 / Vox2 / Libri & 0.1387 & 3.53 & 0.1395 & 3.58  \\
\addlinespace[.3em]
13 & \multirow{2}{*}{ResNet34} & \multirow{2}{*}{A-Prototypical} & \multirow{2}{*}{SAP} & Vox2 & 0.1762 & 3.99 & 0.1769 & 3.96 \\
14 &  &  &  & Deep / Vox1 / Vox2 & 0.1647 & 3.83 & 0.1654 & 3.85 \\
\addlinespace[.3em]
15 & ResNet34 & GE2E & SAP & Deep / Vox1 / Vox2 & 0.1768 & 4.08 & 0.1778 & 4.07 \\
\arrayrulecolor{black!30}\specialrule{.01em}{.15em}{.15em}
16 & \multicolumn{4}{c}{x-vector baseline \textit{(provided by SdSV)}} & 0.5290 & 9.05 & 0.5287 & 9.05 \\
\addlinespace[.3em]
17 & \multicolumn{4}{c}{i-vector/HMM baseline \textit{(provided by SdSV)}} & 0.1472 & 3.47 & 0.1464 & 3.49 \\
\arrayrulecolor{black!30}\specialrule{.01em}{.15em}{.15em}
18 & \multicolumn{4}{c}{Fusion of TDNNs [1-7]$^{\dag}$} & 0.1242 & 3.50 & 0.1257 & 3.55 \\
\addlinespace[.3em]
19 & \multicolumn{4}{c}{Fusion of ResNet34s [8-14]$^{\dag}$} & 0.0940 & 2.40 & 0.0942 & 2.42 \\
\addlinespace[.3em]
\textbf{20} & \multicolumn{4}{c}{\textbf{Fusion of all systems [1-14]}$^{\dag}$} & \textbf{0.0771} & \textbf{2.18} & \textbf{0.0785} & \textbf{2.23} \\
\arrayrulecolor{black}\bottomrule
\end{tabular}%
}}
\end{table*}
%
%
\section{Experimental conditions and Analysis}
\subsection{Training condition}
According to the fixed training condition of the challenge, we used the designated datasets for training our systems and utilized RSR2015 dataset \cite{larcher2014rsr} as a validation set for monitoring. The training set for each system was the combination of different datasets and each training dataset is described as follows.

\vspace{2.0mm}\noindent\textbf{DeepMine (Task 1 Train Partition)}. This is the main dataset, i.e., in-domain data, of the SdSV Challenge. It contains 101,063 utterances from 963 speakers, which have five Persian and five English phrases. We used it for training (1) speaker embedding network and (2) estimating the posterior of phrase, and also as (3) cohort set to calculate parameters of AS-Norm.

\vspace{2.0mm}\noindent\textbf{VoxCeleb1 \& 2}. We used the development sets of VoxCeleb1 \cite{nagrani2017voxceleb} and VoxCeleb2 \cite{chung2018voxceleb2}, which consist of 148,642 and 1,092,009 utterances from 1,211 and 5,994 speakers respectively. In our systems, they were used to train the speaker embedding networks.

\vspace{2.0mm}\noindent\textbf{LibriSpeech}. To train the CTC-based ASR model, namely \(Jasper\), which was utilized in character-level pooling and for estimating the posterior of phrase, we used the train-clean/other sets of LibriSpeech corpus \cite{panayotov2015librispeech}, which comprise 281,241 utterances from 2,338 speakers. Additionally, in some systems, we employed them for training speaker embedding networks.

\subsection{Trial condition}
According to the trial condition of the challenge, the enrollment was accomplished using three utterances of a specific phrase for each model and among four types of trials in the TD-SV task, only Target-Correct, where the target speaker utters the correct pass-phrase, was considered as target and the rest was an imposter. The trial set was divided into two subsets: a progress subset (30\%), and an evaluation subset (70\%). The progress subset was used to monitor progress on the leaderboard, while the evaluation subset was used for the official results.

\subsection{Analysis}
We analyzed two experimental scenarios. First, we verified the feasibility and effectiveness of the character-level pooling strategy for the TD-SV task through the results on the progress and evaluation subsets (Table 3). In the second experiment, we applied AS-Norm and score compensation to all the systems we used in the first experiment, to further boost the performance. Also, we fused different systems and confirmed the best primary system and single system on the progress and evaluation subsets in terms of MinDCF, which was the main metric for the challenge (Table 4). No preprocessing such as data augmentation or VAD was applied to the training and trial data.

\vspace{2.0mm}\noindent\textbf{Analysis of character-level pooling strategy (Table 3)}. Each subsystem (1-15) was composed of different front-ends, pooling techniques, objectives, and training datasets described in Section 2 and 3.1. Among them, system (5) utilized text-dependent training (TDT), which was jointly trained by combined classes of speaker and phrase (i.e., speakers \(\times\) phrases classes). In systems of (5-16), phrase information was not considered, since they were trained for TI-SV. For the reasons stated in Section 2.2, the character-level pooling methods (systems 1-4) showed improved performances compared with other systems in terms of 0.3554 MinDCF and 8.48\% EER on the challenge's evaluation subsets (See Table 3).

\vspace{2.0mm}\noindent\textbf{Results using AS-Norm \& score compensation (Table 4)}. We used AS-Norm and score compensation described in Sections 2.6 and 2.7 respectively, for improvement of performances. As you can see in Table 4, the performance of all systems was improved significantly. In particular, the performances of systems that didn't consider the lexical context (5-15) increased greatly, compared to the character-level pooling systems (1-4), which showed minor improvement. The best performance of a single system was 0.1307 MinDCF and 3.18\% EER on the evaluation subsets. From these results, we could interpret that the better the speaker is distinguished, the higher the performance can be achieved, when score compensation was applied. Finally, we performed the fusion by computing the equal-weighted sum of the scores of different systems, which were TDNN-based (18), ResNet-based (19), and all systems (20). Overall, the performance of ResNet-based systems outperformed TDNN-based systems in the case of both single systems and fusions, and the best primary system was the fusion of all systems. It obtained 0.0785 MinDCF and 2.23\% EER on the evaluation subset.

\section{Conclusions}
In this paper, we described our submission to Task 1 of the SdSV Challenge 2020. We propose a new pooling and score compensation methods that leverage a CTC-based end-to-end ASR model for taking the lexical content into account. Our systems contained two front-end architectures and acoustic features, and various pooling methods including our proposal, and different objective functions.
Experiments show that the usage of the proposed character-level pooling and score compensation methods significantly enhances text-dependent speaker verification performance. Finally, the best performance of the primary system was obtained through the fusion of all the experimented systems, which showed 0.0785\% MinDCF and 2.23\% EER on the challenge's evaluation subset.

\section{Acknowledgements}
This work was supported by the research fund of Signal Intelligence Research Center supervised by Defense Acquisition Program Administration and Agency for Defense Development of Korea.

\bibliographystyle{IEEEtran}
\bibliography{mybib}

\end{document}